\documentclass[10pt,journal,compsoc]{IEEEtran}

\IEEEoverridecommandlockouts                              

\usepackage{cite}
\usepackage{amsmath,amssymb,amsfonts}
\usepackage{algorithmic}
\usepackage{graphicx}
\usepackage{textcomp}
\usepackage{url}
\usepackage{hyperref} 
\usepackage{xcolor}

\hypersetup{
     colorlinks=true,
     linkcolor=black,
     filecolor=black,
     citecolor = black,      
     urlcolor=black,
     }

\title{An Optimized and Safety-aware Maintenance Framework: A Case Study on Aircraft Engine}

\author{Muhammad Ziyad$^{1}$, Kenrick Tjandra$^{2}$, Zulvah$^{3}$, Mushonnifun Faiz Sugihartanto$^{4}$, Mansur Arief$^{2}$%

\thanks{$^{1}$Muhammad Ziyad is with Sustainable Production Development,  KTH Royal Institute of Technology, Södertälje, Sweden}%
\thanks{$^{2}$ Kenrick Tjandra and Mansur M. Arief are with the Department of Mechanical Engineering, Carnegie Mellon University,  USA}%
\thanks{$^{3}$Zulvah is with Aquaculture Industry, Tangerang, Indonesia}%
\thanks{$^{4}$Mushonnifun Faiz Sugihartanto is with the Department of Business Management, Institut Teknologi Sepuluh Nopember, Surabaya, Indonesia}%
\thanks{}%
\thanks{The code for this paper is made available at \url{https://github.com/mansurarief/optimized-safety-aware-aircraft-maintenance}}%
}

\begin{document}

\maketitle
\thispagestyle{empty}
\pagestyle{empty}

\begin{abstract}
The COVID-19 pandemic has recently exacerbated the fierce competition in the transportation businesses. The airline industry took one of the biggest hits as the closure of international borders forced aircraft operators to suspend their international routes, keeping aircraft on the ground without generating revenues while at the same time still requiring adequate maintenance. To maintain their operational sustainability, finding a good balance between cost reductions measure and safety standards fulfillment, including its maintenance procedure, becomes critical. This paper proposes an AI-assisted predictive maintenance scheme that synthesizes prognostics modeling and simulation-based optimization to help airlines decide their optimal engine maintenance approach. The proposed method enables airlines to utilize their diagnostics measurements and operational settings to design a more customized maintenance strategy that takes engine operations conditions into account. Our numerical experiments on the proposed approach resulted in significant cost savings without compromising the safety standards. The experiments also show that maintenance strategies tailored to the failure mode and operational settings (that our framework enables) yield 13\% more cost savings than generic optimal maintenance strategies. The generality of our proposed framework allows the extension to other intelligent, safety-critical transportation systems. 
\end{abstract}


\section{Introduction}
Intelligent transportation systems have undeniably shown rapid development in recent years\cite{smartcities3020018}. Its existence has benefited humans in multiple aspects of our daily lives, e.g. through the public bus tracking system, assisted parking technology, or highway collision avoidance system \cite{hamilton2021hist}. In the coming years, implementing more advanced intelligent transportation systems will put heavier reliance on these systems \cite{9564580} and less intervention or dependency on human aspects, requiring each system to maintain its reliability, often through routine inspection and maintenance procedures. The heavily regulated air transportation industry serves as an excellent example of designing a safety-critical deployment for an intelligent transportation system \cite{Kahraman2022}.

Over the last few decades, airline industries have seen rapid growth \cite{numanouglu2020analysis}. Statistics show that pre-COVID-19 pandemic, the number of passengers traveling by airplanes has increased significantly, from around 1.95 billion people in 2004 to almost 4.6 billion in 2019 \cite{mazareanu2020number}. This huge market potential has attracted numerous airline companies, resulting in unprecedented fierce competition and forcing them to operate as efficiently as possible to gain competitiveness while maintaining compliance with safety standards and regulations and delivering quality services. Since 2020, COVID-19 has exacerbated the situation, causing the airline revenues to plummet to only 40\% of the previous year's value \cite{bouwer2021back}. The significantly heavier pressure forced companies to find means to reduce their operational costs to stay viable in the market, including adopting artificial intelligence (AI) in assisting decision-making process data \cite{xu2021machine}.

One of the components of operational costs that are often put under scrutiny for potential expenses reduction is maintenance \cite{fritzsche2014optimal}. The total aircraft maintenance costs contribute on average 10\%-15\% of the total operational costs \cite{sprong2020deployment}, with a value reaching half of billion dollars. For instance, in 2019, the maintenance costs of full-service airlines such as Qatar Airways, Garuda Indonesia, KLM amounted to 10\%, 12\%, 14\% of the total operating costs, respectively \cite{qatar2019cost, garuda2019cost, klm2019cost}. For low-cost airlines such as AirAsia Berhad, the percentage of maintenance cost is around 11\% in 2019 \cite{airasia2019cost}.  
If optimized, maintenance strategy could deliver appealing cost reductions \cite{rengasamy2018deep}. However, too much inclination on cost reductions could increase the safety risk of passengers and crews, especially if the maintenance standard is compromised to the bottom margin of the minimum safety standard. Therefore, the decision-makers should be equipped with decision-making tools to weigh safety considerations properly.

Despite being among the easiest to implement, preventive maintenance strategy is prone to underestimation and overestimation without systematic assessment. Several drawbacks, for example, are (potentially unnecessarily) high maintenance cost, human error or excessive operational usage, which accelerate engine(s) deterioration, and longer maintenance time in total \cite{horner1997building, pecht2009prognostics}. To get ahead of these barriers, some manufacturers are moving toward more advanced prognosis and health management of the aircraft engine to predict the failure. The advent of sensor measurements provides relatively cheap and fast means to collect signals on the current airworthiness of the engines. The availability of these data enables the prognostics approach and has led to the advancement of more accurate prognostics based on data analytics, machine learning, and AI \cite{chukwuekwe2016reliable, safaei2019premature, xu2020aircraft, xu2021machine}.

In this paper, we leverage on data-driven nature of prognostics to make the first leap in building an AI-assisted framework for an optimized maintenance decision. Our goal is to make safe decisions and use sensor measurements to estimate the hazard score of aircraft engines over time, which serves as a proxy for the safety level of the engine. We then use this score to determine the airworthiness of the engines. Specifically, we use the Cox Proportional-Hazards regression model for our framework and employ simulations to account for the uncertainty of optimal threshold selection in minimizing the total maintenance costs  (performance restoration costs for preventive approach or LLP replacement costs for reactive approach). This threshold can then be used as a criterion to assess the operability of an aircraft, striking a balance between preventive and reactive maintenance measures. Since the problem can be identified in advance,  maintenance cost, lead time, and duration can be minimized, and assets utilization can be maximized.

Our contribution is threefold. First, we advance the use of prognostics for optimizing engine maintenance decisions under uncertainty. By synthesizing the power of proportional hazard model and versatility of simulation-based optimization method, using NASA Turbofan public dataset \cite{saxena2008turbofan} we demonstrate an optimized maintenance strategy in the numerical experiments that attains 37\%-50\% cost reduction compared to a baseline model that could schedule maintenance either overly conservatively or extremely dangerously. Second, our framework allows the decision-makers to construct strategies based on their priority on safety and cost, an important feature to make cognizant maintenance decisions \cite{hopp1998optimal}. Finally, the data-driven nature of our approach enables a more directed maintenance program based on the characteristics of the engines, rather than a simple and generic one that often be used when lacking proper prognostics. We believe that the use of more advanced prognostics techniques that AI promises in the near future will further enhance the robustness of our framework and scale up its real-world applications, including connected and autonomous vehicles.

\section{Related Works}
\label{sec:literature_review}

In this section, we briefly review the concept of aircraft engine maintenance, data-driven prognostics approaches, and optimal maintenance models.

\subsection{Aircraft Engine Maintenance}

The maintenance approach for turbofan engines can be classified into two major categories (or the combinations thereof): scheduled (or preventive) maintenance and unscheduled (reactive) maintenance. Most scheduled maintenance approaches are based on the modules, and the LLP installed on the engines  \cite{shannonackert2020isat}. The scheduled maintenance for turbofan engine is further divided into engine modules performance restoration, Life Limited Parts (LLP) replacement \cite{shannonackert2020isat}, and engine re-delivery program \cite{shannonackert2014lease}. Moreover, engine maintenance accounts for up to 80\% of the total aircraft maintenance costs \cite{shannonackert2020isat}.
 One popular LLP-based approach is the engine removal forecast. This method uses some parameters such as flight cycles or flight hours and advocates that the turbofan engine should be maintained or replaced in the workshop when the LLP reaches its limit. The scheduling often gets highly complicated as each engine consists of many different types of LLPs. Engine re-delivery program is another type of scheduled maintenance program. This approach aims to restore the turbofan engine condition to its best condition by repairing or replacing modules and LLP before being delivered back to the engine lessor. This type of maintenance occurs at the end of the engine operation/usage phase.

In contrast, reactive maintenance is unforeseen in nature. It could occur due to various reasons such as hard landing, ground damage, tail strikes, lightning strikes, bird strikes, or high engine temperature  \cite{shannonackert2018monitor}. Due to the uncertainty of these events, it is difficult to predict their occurrence, the number of resources needed, and the costs to allocate. Frequently, the maintenance events are viewed as a combination of both scheduled and unscheduled maintenance.

The engine maintenance event can be defined based on the bathtub curve \cite{shannonackert2020isat, pecht2009prognostics}. Based on the best practices, the proportion of maintenance is subject to what stage the engine is in its life cycle. In the preliminary stage, the scheduled maintenance event is around 30\%, and the unscheduled maintenance event is around 70\% from the total event. After the engine reaches the mature stage, scheduled maintenance takes a higher proportion than the unscheduled maintenance event (around 70\% for scheduled maintenance and 30\% unscheduled) as its failure rate is lower in the mature stage. In the wear-out stage, the failure rate of the engine tends to increase as the engine age increases. The unscheduled maintenance occurs more frequently at this stage. Thus, the proportion of scheduled maintenance is around 30\%, and the unscheduled maintenance is around 70\% \cite{shannonackert2020isat, pecht2009prognostics}.
By optimizing the maintenance event, the airlines could minimize their maintenance cost and maintain their customers' safety level.

\subsection{Data-driven Prognostics Approach and Survival Analysis}

Prognostics approaches have led to the advancement of predictive maintenance strategies using data analytics and machine learning techniques prevalent success in minimizing the risk of failure and improve efficiencies in many areas including healthcare and engineering  \cite{katzman2018deepsurv, zheng2020condition}. At a high level, prognostics is an effort to predict the remaining lifetime or the failure event using the signals obtained from the measurements. At its core, prognostics uses historical data to find the interrelation between parameters, extract the essential pattern and predict the future state of the system under study \cite{baraldi2012kalman, niu2010development, jardine2006review, chen2012technical, millar2009defining, esperon2013review}. Recent work uses a more complex AI system to improve the prediction accuracy \cite{xu2021machine}. While many of these works have been able to predict the remaining useful life of the engines with sufficient accuracy, using the predictions to formulate an optimal maintenance decision remains a challenge, especially when we are interested in accounting for the prediction error.

A more straightforward prognostics approach is to estimate the airworthiness of the engine based on sensor measurements to model the failure events. A useful technique is called survival analysis. Survival analysis deals with problems with a significant emphasis on the time until the occurrence of a particular event (classical examples include death, breakdowns, failures, etc., hence the term ``survival'') \cite{fox2002cox}.

\subsection{Optimal Maintenance Models}
Maintenance decisions have been among the classical optimization techniques applications with tremendous success history. The literature on the use of mathematical modelling to analyze, plan, and optimize maintenance operations in various industrial contexts is abundant \cite{dekker1996applications, alaswad2017review, vilarinho2017preventive}. The approaches are mainly classified into two: fixed-time and condition-based models, with a good review provided in \cite{ahmad2012overview, shin2015condition}.  We are mainly interested in the later models. Many approach employs linear programming formulations to identify the optimal solutions \cite{ braglia2013integer, almgren2008optimization}. The strength of these workers includes the characterization of the solutions, which often offers valuable insights on how to formulate even higher-level strategies for the companies. However, these approaches are designed based on a prior understanding of the nature of the failures, which could be expensive or even impractical to obtain in real life. Hence, we start with a simulation-based approach to optimize our hazard threshold to be as generic as possible. This choice also allows us to quantify and output the uncertainty of our approach, which can be useful to account for by the decision-makers.

\section{Metholodogy}
\label{sec:methodology}

Our framework mainly consists of two components: engine hazard estimation using Cox Proportional-Hazards model and hazard threshold simulation

\subsection{Hazard Estimation via Cox-Proportional Hazards Model}

In this work, we adopt the Cox Proportional-Hazards model, which is a semi-parametric approach to determine the effect of covariates $x_t \in \mathbb R^d$ on the survival at time $t$. The hazard function for this model is given by
\begin{equation}
    h(t|x_t) = h_0(t) \hat h(x_t),
\end{equation}
which consists of time-varying baseline hazard rate $h_0(t)$ and partial hazard $\hat h(x_t)$ defined as
\begin{equation}
    \hat h(x_t) = \exp \left( \beta^\intercal (x_t-\bar x_t) \right),
\end{equation}
 that signifies the effect of the covariates $x_t$. Here $\bar x_t$ denotes the mean of $x_t$ (computed during training). The parameters $\beta \in \mathbb R^d$ and the baseline hazard $h_0(t)$ is fitted using $\mathtt{lifelines}$ package \cite{davisonpillon2015lifelines}.

 \subsection{Hazard Threshold Optimization via Simulation}
 
 The survival analysis allows us to estimate the current hazard level $\log \hat h(x_t)$ if we can collect sensor measurements $x_t$ at time $t$. With this hazard level estimate as a proxy, we can set a threshold as a decision criterion, say $\lambda$, to determine the airworthiness of the aircraft engines at time $t$. A straightforward binary decision $y_t(\lambda) \in \{0, 1\}$ at time $t$ (where 0 means no maintenance required and 1 means otherwise) could therefore be
 \begin{equation}
     y_t(\lambda) = \begin{cases}
     1, & \text{if }\log \hat h(x_t) < \lambda, \\
     0, & \text{otherwise}.
     \end{cases}
 \end{equation}
 This means we allow the aircraft to fly at time $t$ if the preflight hazard level $\log \hat h(x_t)$ is below our threshold $\lambda$. Otherwise, we cancel the flight and perform maintenance.
 
 We note, however, that our estimate can be very noisy thus needs post-processing. We use a set of training data $\mathcal D_{\text{train}} = \{x_{[1:t_i]}, t_i\}_{i=1}^n$ that contains $n$ engines to optimize $\lambda$. For engine $i$, we assume to have $t_i$ measurements $x_{[1:t_i]} = [x_1, x_2, \cdots, x_{t_i}]$, i.e. the sensor measurements from time $t=1$ to $t=t_i$ after which the engine fails. We use this dataset to simulate and compute the optimal threshold $\lambda^*$ that balances safety and cost. 
 
 For simplicity, suppose that from each flight, the airline incurs $C_1$ engine performance restoration cost if performing preventive maintenance and $C_2 > C_1$ engine LLP replacement cost if the engine fails (all assumed a fixed constant). Then, for a threshold $\lambda$, the total maintenance cost can be written as
 \begin{align}
     \text{TotalCost} (\lambda) &=      
     \sum_{i=1}^n  C_1 \left(\max_{t=1, \cdots, t_i} \{\log \hat h(x_t)\} \geq \lambda  \right) \nonumber \\
     &+ \sum_{i=1}^n C_2 \left(\max_{t=1, \cdots, t_i} \{\log \hat h(x_t)\} < \lambda  \right) \label{eq:total_cost}. 
 \end{align}
 The first term in the RHS of (\ref{eq:total_cost}) shows that restoration cost is incurred if the hazard score is above the threshold. The second term highlights that replacement cost is applied if no maintenance is done because the engine eventually fails at time $t_i$. The optimal threshold $\lambda^*$ is obtained by minimizing the total maintenance cost
 \begin{equation}
     \lambda^* = \arg \min_{\lambda} \text{TotalCost}(\lambda).
     \label{eq:opt_threshold}
 \end{equation}
 We then use the optimized $\lambda^*$ and evaluate the resulting maintenance strategies using data in the test set $\mathcal D_{\text{test}}$.
 
 In addition to the total maintenance cost, we also output the estimated probability of engine failures given $\lambda$:
 \begin{equation}
     \text{FailureProb}(\lambda) = \frac{1}{n} \sum_{i=1}^n \mathbb I\left(\max_{t=1, \cdots, t_i} \{\log \hat h(x_t)\} < \lambda  \right),
     \label{eq:failure_prob}
 \end{equation}
where $\mathbb I(\cdot)$ is an indicator function that outputs 1 if the supplied argument is true and outputs 0 otherwise. Since this probability describes the notion of safety, we need to account for the uncertainty in our estimation.
 
To do this, we employ a simulation-based optimization approach. Instead of using the whole $n$ samples, we sample with replacement $n_0 < n$ engines for $k$ replications and compute the means and standard deviations of the total cost (\ref{eq:total_cost}) and failure probability (\ref{eq:failure_prob}).  The approach helps provide additional robustness to the framework, especially when the estimated hazard scores are very noisy. Finally, we note that while we use constant values for $C_1$ and $C_2$ across time and engines, the approach can directly be extended for when these values are variable.

\section{Numerical Experiments}
\label{sec:experiment}

We use NASA Turbofan engine datasets in our experiment. The datasets are provided by NASA Ames Research Center. These datasets are simulation results generated using simulation software known as Commercial Modular Aero-Propulsion System Simulation (C-MAPSS). C-MAPSS simulates turbofan engines model with the engine specification of 90,000 lb thrust class, operating altitudes between sea level to 40,000 ft, Mach number ranging from 0 to 0.90, and sea-level temperature ranging from -60$^\circ$ to 103$^\circ$ F. Besides, the power management systems can be operated over a wide range of thrust levels across a full range of flight \cite{saxena2008turbofan}.

The datasets are divided into the training set and the test set. There are six combinations of operating conditions that consider the altitude, TRA, and Mach number. The fault modes considered include single fault mode (HPC fault) and multiple fault modes (HPC and fan fault). In total, there are 708 data points (combined), which are then partitioned into four subsets (FD001, FD002, FD003, and FD004).

The turbofan engine datasets were used to construct a model that can minimize the maintenance cost while maintaining the company's safety level . In the experiment, we use the following parameters:  performance restoration cost $C_1 = 3.5 \times 10^6$, LLP replacement cost $C_2 = 4 \times 10^6$, number of sampled engines $n_0 = 30$, and simulation replications $k=10$. We enumerate a range of $\lambda$ values and simulate the system using the training set to obtain $\lambda^*$.

\begin{figure*}[h]
    \centering
    \includegraphics[width=0.5\linewidth]{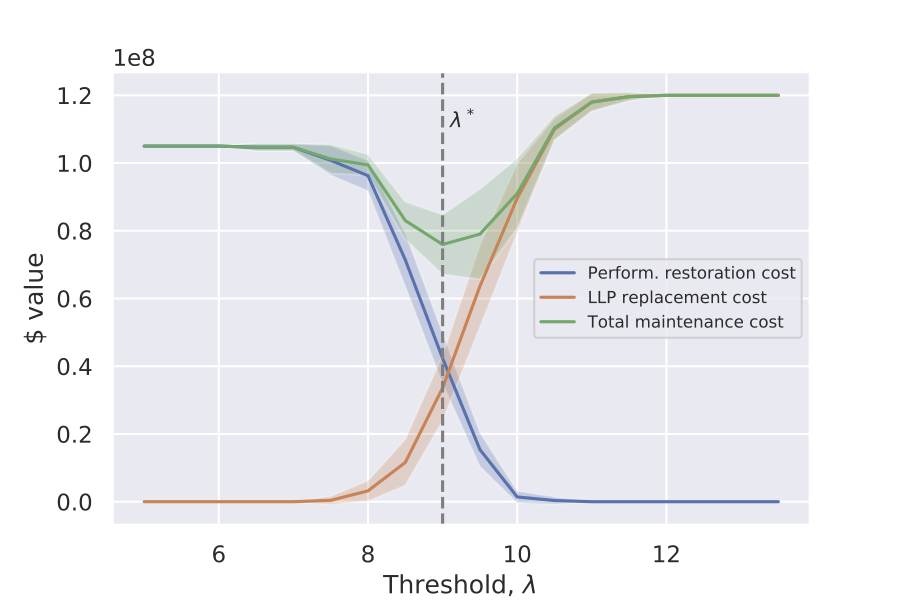}
    \caption{Maintenance cost vs threshold $\lambda$ for the combined dataset. The optimal threshold balances both components of the total maintenance cost.}
    \label{fig:opt_mach_fd00x}
\end{figure*}

\begin{figure}[h]
    \centering
    \includegraphics[width=0.9\linewidth]{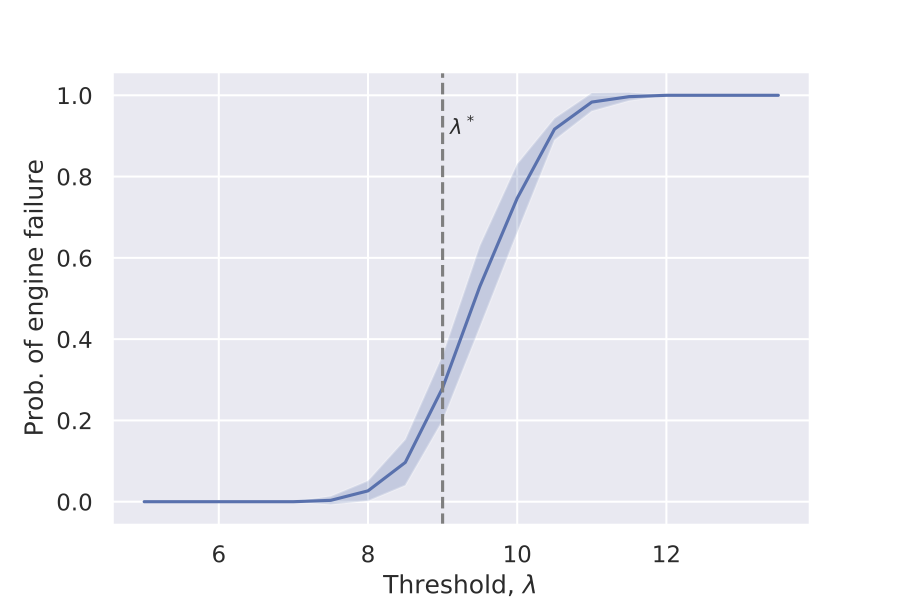}
    \caption{Failure probability vs hazard threshold $\lambda$ for the combined dataset. The optimal threshold that minimizes the total maintenance cost takes a certain level of risk of failure.}
    \label{fig:prob_fd00x}
\end{figure}

\begin{figure}[h]
    \centering
    \includegraphics[width=0.9\linewidth]{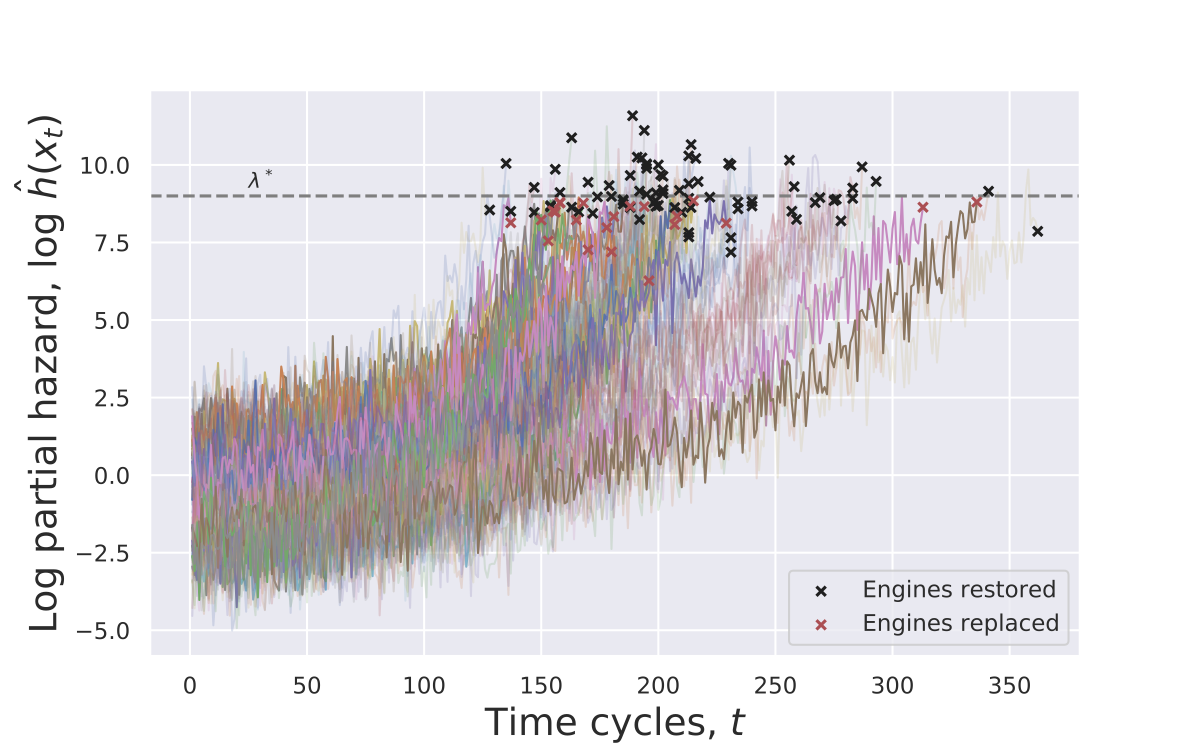}
    \caption{Simulated hazard trajectories for combined dataset. The risk level corresponding to the optimal threshold prevents most engines to break down and leaves only a few to be unmaintained (hence gets broken and replaced).}
    \label{fig:sim_result_fd00x}
\end{figure}

We first implement the proposed method to obtain the optimized threshold for all engines (combined dataset), giving us the generic maintenance strategy for all operating conditions.  The costs for various $\lambda$ values are summarized in Fig. \ref{fig:opt_mach_fd00x} while the probabilities are shown in Fig. \ref{fig:prob_fd00x}. The colored lines in these figures represent the means for the costs and failure probability, while the shadows represent their one-sigma confidence intervals, i.e. the interval  $[\text{mean}-\text{std}, \text{mean}+\text{std}]$. Our simulations show that this choice of $\lambda^*$ gives about $0.28$ probability of engine failures, with simulated hazard trajectories shown in Fig. \ref{fig:sim_result_fd00x}.

\begin{figure*}[ht]
    \centering
    \includegraphics[width=0.8\linewidth]{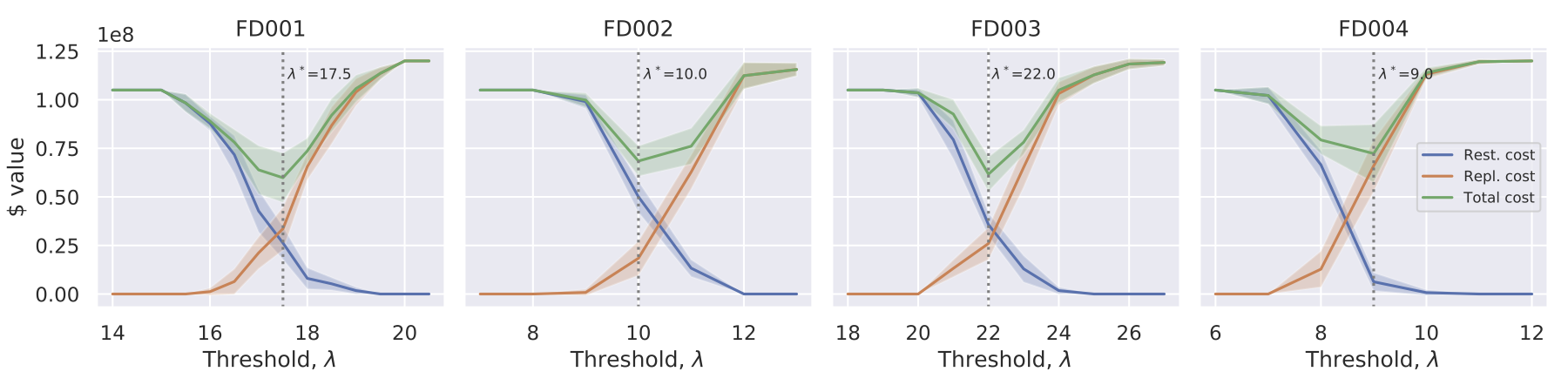}
    \caption{Maintenance cost vs hazard threshold $\lambda$ for FD001, FD002, FD003, and FD004. The optimal threshold varies among datasets, suggesting the use of more directed maintenance decisions could lead to higher cost reductions.}
    \label{fig:fdall_cost}
\end{figure*}

\begin{figure*}[ht]
    \centering
    \includegraphics[width=0.8\linewidth]{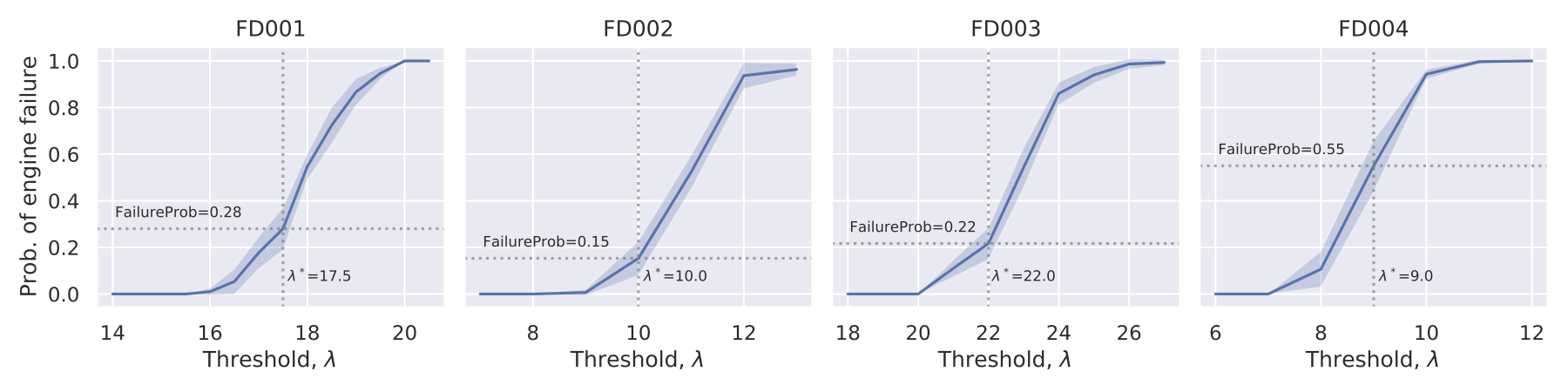}
    \caption{Probability of engine failure vs hazard threshold $\lambda$ for FD001, FD002, FD003, and FD004. The probability of engine failure highly varies based on the optimized threshold for each dataset. The cost-optimized threshold leads FD004 with the highest hazard score variability to take extremely high risks. Decision-makers should pay attention to this and might want to correct and lower the threshold to avoid taking such a risky strategy.}
    \label{fig:fdall_prob}
\end{figure*}

Finally, we implement the proposed approach to all four partitions of the datasets (FD001, FD002, FD003, and FD004). We then compare the costs and safety (in terms of probability of engine failures) for the optimized threshold. The results are summarized in Fig. \ref{fig:fdall_cost} and \ref{fig:fdall_prob}.

\section{Discussion}
\label{sec:discussion}

In this section, we discuss our findings. First, Fig. \ref{fig:opt_mach_fd00x} shows that the optimal cost for engine maintenance for the combined turbofan engine dataset is achieved with $\lambda^* = 9.0$. When the hazard score (log-partial hazard) exceeds this threshold, the engines are likely to fail due to HPC or fan faults. If it happens, the failure event would then accrue more maintenance costs to the airlines, leading to higher LLP replacement costs. The optimal maintenance cost for the combined turbofan dataset estimated at around US\$ 76 Million. In this case, we can reduce the maintenance costs  37\% under the optimized maintenance strategy.

Next, we highlight in Fig. \ref{fig:prob_fd00x} the failure probability on different failure threshold. While $\lambda^* = 9.0$ indeed minimize the cost, the decision-makers could adjust this threshold to account for his safety priority. For instance, if the current engine failure probability ($\text{FailureProbability}(\lambda^*) = 0.28$) is deemed too high, thus too risky, then a lower $\lambda$ values could be chosen. From Fig. \ref{fig:sim_result_fd00x}, one shall see that doing so will increase maintenance frequency, thus more engines will be restored, and less will be left broken and replaced since the threshold for the hazard will be lower. This setting, in turn, will increase safety (but comes with higher maintenance costs accordingly). This sort of analysis is one of our approach's advantages: it empowers decision-makers with quantitative measures to weigh safety and efficiency considerations carefully.

In Fig. \ref{fig:fdall_cost}, we show the optimal cost for engine maintenance for FD001, FD002, FD003, and FD004  turbofan engine dataset. The optimal thresholds w.r.t. minimum maintenance costs are attained at 17.5, 10, 22, and 9 for FD001, FD002, FD003, and FD004, respectively. From the figure, we see that in the case when $\lambda < \lambda^*$, performance restoration costs dominate the maintenance cost. In this case, maintenance is performed too excessively. On the contrary, when $\lambda > \lambda^*$, the replacement costs dominate, suggesting that maintenance is too rarely scheduled, causing far too many engines to be broken and have to be replaced. The optimized threshold $\lambda^*$ balances the two costs. In these cases, we see that the maintenance costs can be reduced by 40\%-50\% under the optimized maintenance strategy.

In Fig. \ref{fig:fdall_prob}, we show the probability of engine failures under various $\lambda$ values for FD001, FD002, FD003, and FD004  turbofan engine dataset. The optimal thresholds yield ``profit-optimized'' failure probability of 0.28, 0.15, 0.22, and 0.55 for FD001, FD002, FD003, and FD004 respectively. One might argue that these failure probabilities are perceivably high, especially for FD004, in which the simulations suggest that the estimated hazard scores are very noisy, suggesting risky airline operations. This situation poses significantly more challenges in determining the optimal threshold (partially captured by the wider confidence intervals for the total cost of FD004 in Fig. \ref{fig:fdall_cost}), leading to more than half of the engines failing and thus being replaced. Hence, one could adopt a more safety-aware decision, shifting $\lambda^*$ slightly lower. Although such adaptation leads to higher costs, it provides an additional buffer against noisy estimations and safer maintenance decisions.

Finally, we note that the cost reductions of using a more directed strategy (using clustered datasets, like FD001, FD002, FD003, or FD004) provide higher cost reductions (about 13\% more) compared to a generic strategy (i.e. using a combined dataset). The data-driven approach allows a more targeted strategy depending on the situations described by the data. However,  we note that in the real world scenario, it might be difficult for an airline to classify the engine based on the failure mode and operating conditions a priori. Therefore, a generic strategy that can cover all engines could be initially implemented while the operators look for ways to identify the engine classifications or clusters based on their failure mode or operational settings.

\section{Conclusion}
\label{sec:conclusion}
In this work, we propose an optimized maintenance strategy developed by combining prognostics and simulation-based optimization. The proposed framework uses the Cox Proportional-Hazards regression model to estimate the hazard score of an aircraft engine and simulation to optimize the hazard threshold, minimizing the total maintenance cost while maintaining a certain safety level. Compared to the generic optimal maintenance method, our proposed work yields 13\% cost reductions. The method also benefits decision-makers, enabling them to design a more directed maintenance strategy if the datasets are clustered based on some characteristics and conditions of the engines. More advanced machine learning methods and AI applications will extend the applicability of the proposed framework to deal with more complex maintenance strategies.

\section*{Acknowledgment}
The authors would like to thank the GMF AeroAsia team for providing valuable discussions and insights about maintenance approaches in airline industries. The first author is extremely grateful for the funding provided by the Swedish Institute to study at KTH Royal Institute of Technology.

\bibliographystyle{IEEEtran}
\bibliography{references}

\end{document}